# Optimal phenotypic plasticity in a stochastic environment minimizes the cost/benefit ratio[1]

Patrick Coquillard[1], Alexandre Muzy[2], Francine Diener[3]

(1) Institut Sophia Agrobiotech (TEAPEA), UMR 7254 INRA-CNRS-Université de Nice-Sophia Antipolis, 400 route des Chappes, P.O. box 167, 06903, Sophia-Antipolis, France.

(2) Laboratoire LISA, UMR CNRS 6240 and Laboratoire I3S, UMR CNRS 7271 (Bioinfo Team Université de Nice-Sophia Antipolis), Università di Corsica - Pasquale Paoli, Campus Grossetti, P.O. Box 52, 20250 Corte, France. a.muzy@univ-corse.fr.

(3) Laboratoire de Mathématiques J.-A. Dieudonné UMR CNRS 6621, Université de Nice - Sophia Antipolis, Parc Valrose, 06108 Nice Cedex 02, France. francine.diener@unice.fr

(1)       Correponding author: patrick.coquillard@unice.fr ; phone: #33 4 92 38 64 30

Abstract

This paper addresses the question of optimal phenotypic plasticity as a response to environmental fluctuations while optimizing the cost/benefit ratio, where the cost is energetic expense of plasticity, and benefit is fitness. The dispersion matrix $\Sigma$ of the genes' response (H = ln|$\Sigma$|) is used: (i) in a numerical model as a metric of the phenotypic variance reduction in the course of fitness optimization, then (ii) in an analytical model, in order to optimize parameters under the constraint of limited energy availability. Results lead to speculate that such optimized organisms should maximize their exergy and thus the direct/indirect work they exert on the habitat. It is shown that the optimal cost/benefit ratio belongs to an interval in which differences between individuals should not substantially modify their fitness. Consequently, even in the case of an ideal population, close to the optimal plasticity, a certain level of genetic diversity should be long conserved, and a part, still to be determined, of intra-populations genetic diversity probably stem from environment fluctuations. Species confronted to monotonous factors should be less plastic than vicariant species experiencing heterogeneous environments. Analogies with the MaxEnt algorithm of E.T. Jaynes (1957) are discussed, leading to the conjecture that this method may be applied even in case of multivariate but non multinormal distributions of the responses.

---

# 1. Introduction

Phenotypic plasticity is a general and common feature, probably shared by most organisms. This concept refers to the ability of genetically identical organisms to change their phenotype in response to environmental changes in space and time. These changes are favoured because they reduce the fitness variance from generation to generation. As a matter of fact, the phenotypic plasticity enables individuals of a population to colonize various ecological systems, to extend the geographical area of the species and thus to reduce its probability of extinction. For instance, a plant grown in a sunny habitat will exhibit broader leaves and a shorter stem than a genetically identical plant grown in a shadowed place. This physiological effect, called etiolation, is an adaptive response to a particular state of the environment, *i.e.*, the quality and the intensity of light the plant receives. Etiolation allows the plant shadowed by neighbours competing with them in the struggle for light and to survive in spite of unfavourable conditions.

Environmental changes do not only affect permanently the phenotype produced by the genotype such as morphological and/or life-history traits. They can also induce several reversible variations of the phenotype (non permanent effects) throughout the life of individuals (Lynch and Jones, 1998). Non permanent effects were particularly studied by behavioural ecologists and many studies were focused upon the fact that organisms can adaptively adjust their behaviour to the environmental heterogeneity during their lifetime.

The empirical concept of reaction norm (RN) is usually defined as "the set of phenotypes [including behavioural phenotypes (A/N)] that a single genotype produces in a given set of environments. Genotypes or individuals show phenotypic plasticity if their RN is non-horizontal" (Dingemanse et al., 2009). As shown in fig. 1, RN elevation and slope (plasticity) can vary among genotypes, and it turns out that some species are more plastic than others when confronted with environmental changes (Via et al., 1995; DeWitt et al., 1998). In addition, some genes can considerably vary in their expression level, while others are more or less constant. These differences probably depend on the functions of the genes.

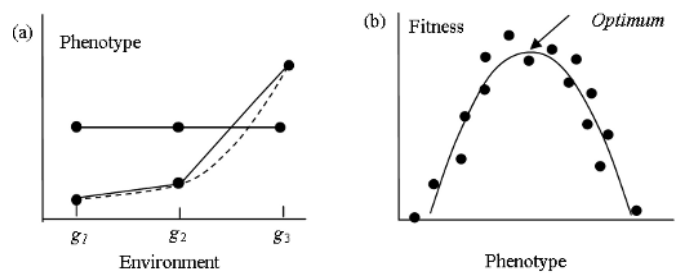

**Figure 1**. Phenotypic plasticity and fitness. (a). Phenotypic plasticity can be described as the amount of change across environmental states. The set of parameters (a vector) of a polynomial regression (dotted line) is used to describe the reaction norm (RN). The horizontal RN depicts a null plasticity (from Via *et al*. 1995 (modified)). (b). The relationship between phenotypic values and fitness shows the limits of plasticity in regard to one environment. A given genotype makes a phenotypic error if its mean phenotype differs from the optimum. The cost of such an error is the difference between the expected fitness and the optimal fitness (from DeWitt *et al*. 1998 (modified)).

The benefit of adaptive plasticity is that organisms can match across more states of the environment in comparison to organisms able to produce a single phenotype only. Currently, it is advocated by several authors that the most plastic phenotypes are harboured by organisms competing in heterogeneous environments relatively to those living in homogenous environments (Ellers and Van Alphen, 1997; Van Kleunen and Fisher, 2005; Weinig, 2000; Snell-Rood et al., 2010).

It is now clear that adaptive phenotypic alters a variety of interactions between individuals and their environment through life-traits and/or behaviour : indirect interactions in multi-species assemblages, direct interactions between species such as inducible defences/offenses (Miner et al., 2005; Ramos-Jiliberto, 2003; Ramos-Jiliberto et al., 2008) and niche construction (Donohue, 2005). Because phenotypic plasticity is able to modify such interactions, it ultimately affects ecological processes, such as population stability, trophic relationships, population dynamics, species coexistence within communities and biodiversity (Mouritsen and Poulin, 2005). Lastly, studies demonstrated that plasticity can promote stability and health of ecological systems experiencing environment fluctuations (Miner et al., 2005 and references therein).

However, organisms able to exhibit a perfect plasticity are unlikely to exist. Such perfect organisms should produce the best phenotype in every condition of the environment. Indeed, most of organisms fail to produce the exact optimal phenotype for various reasons: inability to produce it (due to evolution constraints), inability to get reliable cues on the state of the environment and inability to pay the plasticity cost (Auld et al., 2010). Energetic cost is a constraint which limits the evolution and the development of plasticity. In this respect, DeWitt *et al.* (1998) have listed a set of five main potential costs of phenotypic plasticity. Among them, these authors emphasize the "production cost", which should be considered only if the cost of production by plastic genomes exceeds those for fixed genotypes producing the same phenotype. This means that the variance associated to the average genome expression has an energetic price. The wider the variance, the higher the cost. However, the true nature of the undeniable cost of plasticity is still somewhat obscure and remains an open discussion.

In this paper, we focus on the estimation of the best cost/benefit ratio of plasticity in the context of fluctuant (stochastic) environments. Since the energetic cost paid is positively correlated to the variance of the response to environmental fluctuations, the arising question is: what is the optimal variance of a phenotype undergoing a fluctuating environment which minimizes the cost/benefit ratio?

First, we study the results obtained through the simulations of a model. Three genes are involved in a single phenotypic response driving the behaviour of a virtual wasp, which aims at maximising – by mean of a genetic algorithm – its progeny when confronted to a fluctuating environment. We find that in the course of the optimisation process, according to theoretical works in the domain, the logarithm of the determinant of the dispersion matrix of the three genes is an indicator of the convergence. Second, we build an analytical model of three genes having analogies with the previous one. This model is an attempt of generalizing the previous empirical approach. The optimal response (i.e. the minimisation of the cost/score ratio) is then found using the Lagrange's constrained optimization method, on the basis of a conjecture we made about the energy available and the covariance matrix of the 3 genes response. The analysis of the simulation results inductively corroborates the modelling assumptions of the analytical model and constitutes a qualitative cross validation of both models.

On one hand, results concern some important conjectures already formulated by several authors about the origin of genetic diversity and phenotypic plasticity in various contexts. On the other hand, we argue that the method we used has some analogies with the MaxEnt algorithm (Jaynes, 1957a, 1957b) applied to a multidimensional system and likely provide an estimator of the maximum exergy the system can own in a specific context. However, it does not constitute a rigorous application of E.T Jaynes' method, but rather a transposition of his ideas and some mathematical developments are still in course in order to identify theoretical underpinnings.

## 2. Numerical model

### 2.1. Model description

#### *2.1.1. Animal behaviour*

The model refers to the foraging behaviour of a parasitoid[2] insect (a small wasp). Some details of the algorithm are indicated in Fig. SD1. The model has been conceived as simple and neutral as possible. It is noteworthy that *we did not aim at reproducing the behaviour and particular traits of a particular species* but rather some general traits of thelytokous (*i.e.*, without sexual reproduction), solitary (*i.e.*, laying only one egg per host) and synovigenic females of parasitoid insects whose plasticity is likely a key factor of survival and fitness in fluctuant environments. It is based on the existence of a well documented trade-off between survival and reproduction resulting from competitive allocation of resources to either somatic maintenance or egg production. In synovigenic species females have the ability to mature eggs throughout their life. A dynamic control of egg load enables animals to retain some flexibility during the adult life and to minimize the risk of experiencing time or egg limitation. For simplification, we considered the case of female wasps that have an instantaneous vitellogenic activity. In addition, animals do not feed during the oviposition stage. Consequently, young females start their life with a limited amount of reserves.

At initialization, a new wasp is instantiated. Immediately after the wasp has emerged, the animal (a female) has the task to lay its eggs. To achieve this task, the female has to forage within an environment in which the resource (hosts) it is looking for (*e.g.* eggs from another species) is aggregated into many patches dispersed within an area. The behaviour of the wasp consists in both finding patches and attacking the greatest possible number of hosts. When a wasp encounters a patch of hosts, the cumulative number of hosts it attacks ($N_t$) obeys to a saturation function of time *t*, admitting as parameter the initial number of hosts within the patch ($N_0$):

$$N_t = N_0(1-\exp(-\alpha t)), \qquad (1)$$

where α is a parameter.

Thus, the velocity for attacking hosts decreases in time, following a negative exponential function. According to the Marginal Value Theorem (Charnov, 1976), when its velocity has reached the average velocity calculated on the basis of the average environment richness, the wasp leaves the patch and tries to find a new one. Such an optimal patch-leaving policy was adopted since the goal of the model was to look for optimal reproductive strategies, and because most foraging animals, especially insect parasitoids, are usually behaving in good qualitative agreement with the theorem. The cycle "foraging for hosts on patches and travelling between patches" is repeated until the wasp has reached the end of its life or has exhausted its potential fecundity.

### 2.2.2. Plasticity.

Three processes driving the wasp behaviour are here represented by three real values coded on 32 bit-structures called genes. The wasp starts its life at an initial position along a linear trade-off (coded by gene *G1*) between its lifespan and its fecundity (fig. 2). Max lifetime and max fecundity are arbitrary fixed to 1000. Using different (realistic) values would lead to a change in scale without affecting qualitatively the results obtained. The animal is able to change its reproductive strategy throughout its life through gene *G2*, which defines the limits of its plasticity. The cost linearly reduces both fecundity and lifespan of the animal by means of a linear relationship with the value of *G2*. The wider is *G2*, the heavier will be the cost (in energy). The appropriateness of a non-symmetric plasticity cost was examined, but we decided to implement a symmetric cost so that the effects on the two characteristics are of the same order of magnitude and can be directly compared. In addition, the wasp can move within the range thanks to gene *G3*, which is a parameter of a linear estimator (McNamara and Houston, 1987), evaluated as follows:

---

[2] An organism that lives at the expense of another (its host), impedes its growth and eventually kills it. Insect parasitoids, which are often very tiny, attack a single organism (plant or animal), from which they derive everything they need for their own growth and reproduction. One way a parasitoid does this is by laying its eggs in the body or eggs of the host insect (after Canadian Forest Service glossary).

$$\mu_i = \lambda_i G_3 + (1 - G_3)\mu_{i-1} \tag{2}$$

The animal has an initial estimate of the encounter rate of $\mu_0$ that was fixed to the midpoint between the lowest (*i.e.*, 0.0) and the highest (*i.e.*, $1/t_1$, where $t_1$ is the time to find the first host on a new discovered patch) possible instantaneous host-encounter rate.

The higher $G_3$, the higher the effect of the past experience on the foraging strategy of the wasp. Every time step of its life, the animal uses $\mu_i$ to decide which strategy to use within the range of possible values. Correlatively, the higher the instantaneous estimated host encounter rate $\lambda_i$, the more the animal will invest in fecundity and the less in longevity.

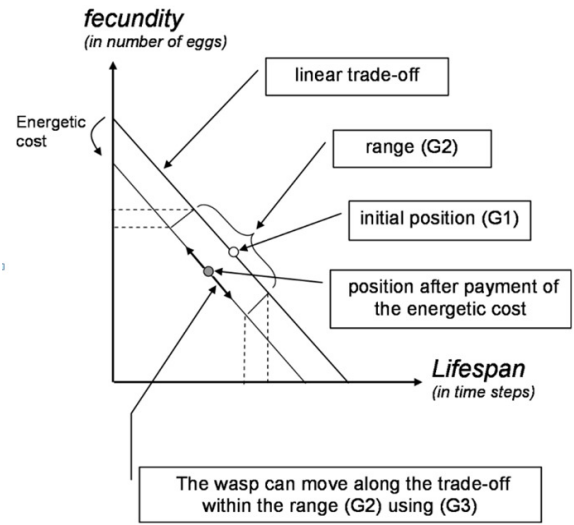

**Figure 2**. The life span-fecundity trade-off of a parasitoïd wasp (Coquillard et al., 2010). G1: initial position of the wasp along the fecundity-lifespan trade-off. G2: Range of the plasticity. The wasp can move within G2 throughout its life to optimize its fitness. G3: a parameter equivalent to sensitivity to the environment richness (Eq.(2)). The energetic cost is the cost of the plasticity which is proportional to G2.

### 2.2.3. Optimization process

Each wasp holds a single chromosome, which encapsulates the three genes. The goal of the simulation consists in finding the vector (*G1, G2, G3*), which maximises the score of the wasp, *i.e.*, the number of eggs laid throughout its life. The score maximization is obtained by means of a genetic algorithm based on the GENITOR model (Whitley, 1989). Crossing-over and mutation rates were fixed to 60% and 2.5% respectively. The population size is 300 and the number of generations 500. The convergence (stability of scores) of the algorithm was carefully verified by runs starting in various initial conditions (average richness of patches, inter-patch travel time, probability to be born on a patch and energetic cost).

### 2.2.4. Environmental fluctuations

The environment in which each simulated animal has to forage is defined by both inter-patch travel time *T* (dilution of patches) and by the average number of hosts on patches ($N_0$, *i.e.*, the patch quality). Each animal of a population should experience the same environment. However, stochasticity is introduced in the model at two levels. First, the wasps can experience environments in which each patch differs from each other by the number of hosts they own. The number of hosts per patch is then drawn from a Gaussian distribution $N \to (N_0 = 100, \sigma^2 = 50)$. Second, the animals can randomly born inside or outside a patch with probabilities *p* and (1-*p*). Thus, animals start their life with or without an inter-patch travel to accomplish. These features aim at simulating the fluctuation levels of the environment. The simulation of each wasp is repeated 20 times, and the score related to each chromosome is obtained by averaging 20 scores. Lastly, 20 replicates of each simulation are performed, each of them differing by the initial values of the populations' chromosomes. The best values obtained in the population are recorded at each time step for each generation and used to compute the dispersion matrix of the genes (see hereafter) over the 20 replicates of the simulation.

## 2.3. Results

A polynomial approach is usually used to describe the RN (*i.e.*, the set of phenotypes) of genotypes confronted to a defined set of environments (Via et al., 1995; De Jong, 1995). In this context, the genetic dispersion matrix **Σ** of character states and the matrix of polynomial coefficients are related by:

$$\mathbf{\Sigma} = \mathbf{XGX}^T, \qquad (3)$$

where the matrix **X** contains a polynomial series of each variable *x* which measures the state of the environment. The relation between the vector of the states of the environments and the coefficients of the polynomial function for a given genotype is then:

$$\mathbf{z} = \mathbf{Xg} \qquad (4)$$

Thus, we retained $\ln[det(\mathbf{\Sigma})]$ as a descriptor of the response variance of the genes confronted to the environment stochasticity in the simulator. For simplicity we will call it the "H value", *i.e.* $H = \ln[det(\mathbf{\Sigma})]$. The results we present hereafter were obtained using a single combination of parameters (maximal plasticity cost, inter-patch travel time = 100, *p* = 0.5) *are representative of the model behaviour* and well suited to test the validity of the H value as a metric of the convergence. A complete analysis of the results is the subject of another paper (Wajnberg et al, 2012). Results show that when the algorithm converges to the best scores, the variance for each gene drastically diminishes but never vanish (fig 3). Indeed, at the end of the simulation there still persists a variance of both responses and scores (table II). This residual variance is interpreted as the effect of the stochastic fluctuations of the environment, which prevents the algorithm converging toward a single and optimal value of the scores.

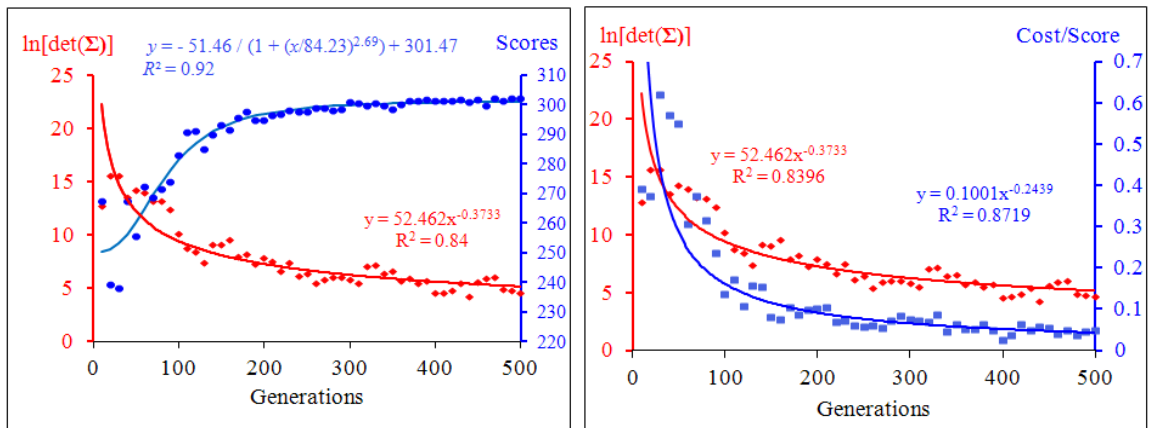

**Figure 3**. Dispersion Matrix and Cost/score ratio. **Left:** convergence of the genetic algorithm to best scores (●) and corresponding H value (♦) of the three genes system. **Right**: In the same time, the Cost/Score ratio diminishes toward an optimum value. Fitted curves are only indicative.

Table I. Mean and standard deviation of genes and scores obtained after the convergence of the genetic algorithm (500[th] generation of the numerical model).

|      | G1     | G2    | G3   | Score  |
|------|--------|-------|------|--------|
| **Mean** | 587.02 | 21.41 | 0.47 | 373.26 |
| **SD**   | 10.94  | 4.81  | 0.11 | 3.89   |

## 3. Analytical model
### 3.1. Model description
By means of an analytical model we now attempt to generalize the results obtained in the previous section. Let $\{G_1, G_2, G_3\}$ be 3 genes that contribute to a phenotype in response to an environmental factor. Individual responses of the genes are modelled by means of a saturation function:

$$y_{Gi}(g(t)) = y_{Gi}^{\max}[1-\exp(-\alpha_i g(t))]^{\beta_i} \tag{5}$$

where, $y_{Gi}$ is the response of the gene $G_i$, $y_{Gi}^{\max}$ is the maximal response of $G_i$, $\alpha_i$ and $\beta_i$ are, respectively, the velocity parameter and shape parameter of $y_{Gi}$ and $g(t)$ is the state of an environmental factor at the $t$ instant. As a consequence, even if several genes share a unique $\alpha$, but different shapes $\beta_i$, the responses will quantitatively differ. This equation is of concave type and is one of the most commonly used to model genes' responses. It results from Eq. (5) that there is no time-lag between experiencing a new signal $g$ and the corresponding phenotype. Similarly, *we are not considering the case of convex functions* since they usually either describe the toxicity effect of some compounds (degradation of various metabolic pathways) or are the result of an inhibitory effect of an epistatic gene. Figure 2a exemplifies the responses of the genes to $g$, each response depending on $\alpha_i$ and $\beta_i$.

In the following, it is considered that the *velocity parameter $\alpha_i$ is homologous to the variance (plasticity) of the gene i response*: $\alpha_i \propto \sigma_i^2$. The equivalence between the parameter $\alpha_i$ and the variance $\sigma_i^2$ of the response is justified by considering that a gene which holds a high variance/plasticity of its response has, on average, a higher response to the environment signal (plasticity in the breadth of adaptation (Gabriel et al., 2005; fig. 4b), whatever the value of the environment, than a gene that can only react to a small range of the environment.

The energetic cost ($E_i$), associated to the $i^{th}$ gene response, according to DeWitt (1998) and Svanbäck *et al.* (2009) is conceived as the sum of several components. Although these authors distinguish five plasticity costs, we aggregated these five items into two categories: the maintenance cost $E_i^{\min}$ (*i.e.*, sensory and regulatory mechanisms) and the cost associated to the variance of the phenotype (production costs, information acquisition costs, developmental instability costs and genetic costs). Svanback *et al.* use a linear function to quantify the energetic cost associated to the variance. Here, an exponential or a power function is preferred, since the variance is a second order moment. Lastly, an additional cost ($\xi$) paid for the response itself is considered. This cost should be equivalent to the cost paid by a fixed genotype to produce an equivalent phenotype. The resulting equation for cost takes the general form:

$$E_i = E_i^{\min} + \gamma_i \exp(\tau_i \alpha_i) + \xi_i y_{Gi}(g(t)), \tag{6}$$

where $\gamma, \tau$ and $\xi$ are positive constants.

The overall energetic cost $E$ is then accounted as the sum of the 3 individual costs $E_i$, over the interval of time $\Delta t$:

$$E_{\Delta t} = \sum_{i,t} E_i(g(t)) \tag{7}$$

We now introduce a constraint on variances, and consequently on costs, that can be understood either as a limitation of the available energy or as an elementary representation of an epistatic link between the genes:

$$\alpha_1 + \alpha_2 + \alpha_3 = Z \quad (Z \geq 0) \tag{8}$$

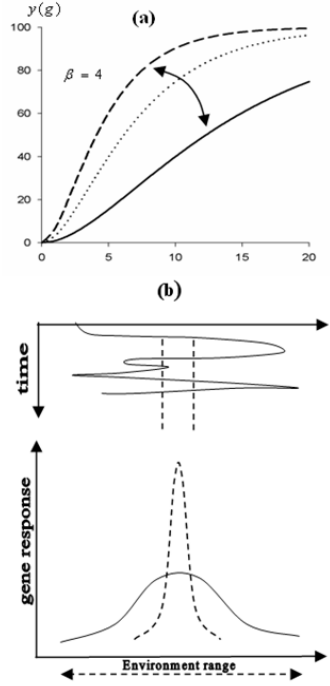

**Figure 4.** Gene response and plasticity. **(a)** Hypothetical example of gene response confined to a limited range ($\beta$ = 4.0). The arrow indicates an example of variation due to the velocity parameter $\alpha$. **(b)** A gene of narrow plasticity will have, on average, a smaller response than a gene owning a wider one, when confronted to a fluctuant environment over a large enough interval of time.

Let us suppose now that the environment is fluctuant through time (*e.g.* for seasonal effects, day/night temperature, inter-annual variation of pluviometry, etc.). The three-genes system is confronted with two kinds of fluctuant environments:

*Periodic fluctuations.*
$$g(t) = a\sin(\omega t + \varphi), \quad \text{where } a \text{ is the amplitude of the variation} \tag{9}$$

*Stochastic fluctuations.*
We tested (i) a uniform distribution (white noise) of same amplitude (*a*) and extremes as (9):
$$g(t) = \min(g) + \left[(\max(g) - \min(g))\right]\mathcal{U}, \tag{10}$$
where $\mathcal{U}$ follows a uniform distribution on [0, 1[, and (ii) a Gaussian distribution (red noise) $N \to (\mu, \sigma^2)$ with parameters drawn from the value of *g(t)* (Eq. (9)) calculated over 100 time steps. Examples are shown on fig. 3.

The response of the three genes is thus a function *S* of the three gene responses, over the interval of time $\Delta t$ (fig. 3):
$$S_{\Delta t} = \sum_{i,t} \partial_i y_i(g(t)) \tag{11}$$

It is assumed that the higher is the response, the higher is the fitness.

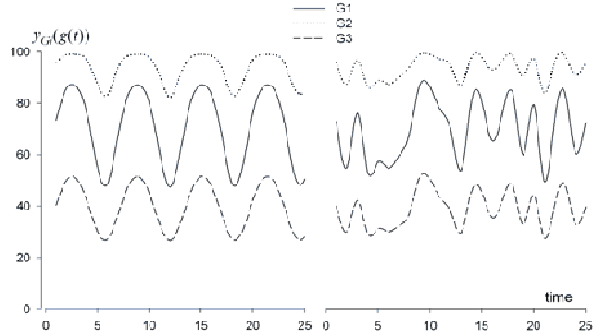

**Figure 5**. Examples of the response of three genes over an interval of time $\Delta t = 25$, submitted to environment fluctuations. Left: periodic environment (Eq. (9)). Right: stochastic environment (Eq. (10)).

It turns out now that the problem consists in finding the best combination $\{\alpha_1, \alpha_2, \alpha_3\}$ which maximizes the response $S_{\Delta t}$, minimizes the cost $E_{\Delta t}$ and satisfies the constraint (8). Notice that there are an infinite number of possible combinations of $\alpha_i$ for each value of Z. To reach this goal, we formulate the following conjecture:

*The phenotypic plasticity, measured by the determinant of the dispersion matrix of genes jointly contributing to the response to environment fluctuations, has its optimum for a particular value Z of the available energy which minimizes the E/S ratio.*

We recall that the determinant of a covariance matrix has the property: $0 < |\Sigma| \leq \prod_{i=1}^{n} \sigma_{ii}^2$ (Hadamard's inequality) and that $\ln|\Sigma|$ is concave (a proof can be found in (Cover and Thomas, 2006); see also fig. SD4). Consequently, we performed the optimisation (maximisation) of $H = \ln|\Sigma|$ subject to the constraint (8), $\{\alpha_1, \alpha_2, \alpha_3\}$ taken as free variables, by means of the standard Lagrange's constrained optimization method.

### 3.2. Results

We partially explored the model: most of the parameters are fixed to neutral values (table II) and the constrained optimization of $H = \ln|\Sigma|$ is performed relaxing the constraint (8), Z varying from 0.2 to 4.

Table II. Parameters' value used for computations

| Parameter | Value | Comments |
|---|---|---|
| a | 2 | Amplitude of the periodic environment |
| $\omega$ | 1 | Phase of the periodic environment |
| $\varphi$ | 0 | Phase shift of the periodic environment |
| $\{\beta_1, \beta_2, \beta_3\}$ | $\{2.0, 1.8, 1\}$ | Shape parameters of genes response |
| Z | $[0.2,…, 4]$ | Constraint on genes; step = 0.2 |
| $E_i^{min}$ | $0, \forall i$ | Same minimal energetic cost for all genes |
| $\xi_i$ | $0.2, \forall i$ | |
| $\gamma_i, \tau_i, \partial_i$ | $1, \forall i$ | No weight |
| $\Delta t$ | 25 | Time Interval |

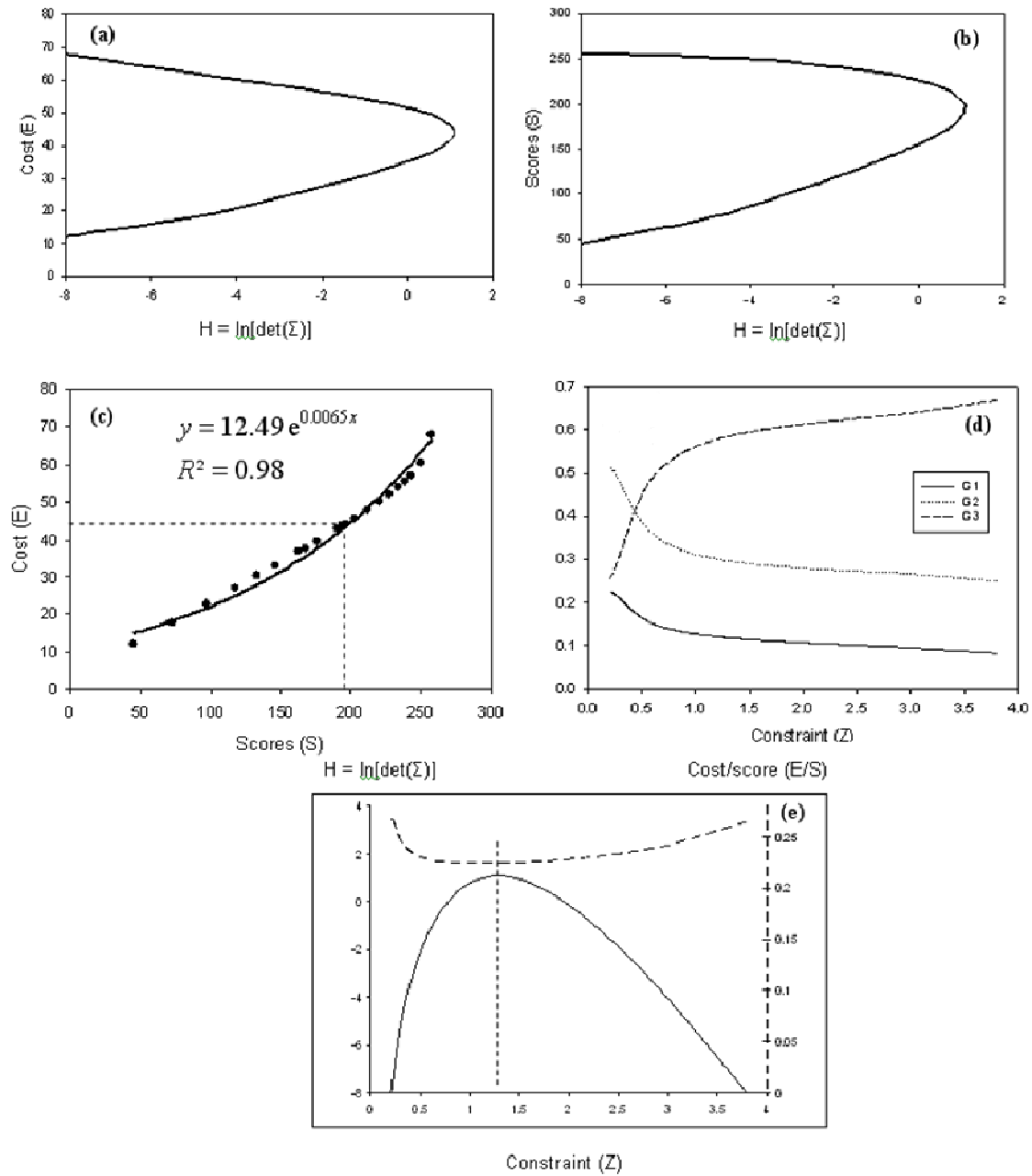

**Figure 6.** Optimal plasticity for the three genes model (example of Gaussian fluctuations of the environment). **(a, b, c)** Optimal cost and score are neither minimal nor maximal, but are intermediate values representing the best compromise. The minimal ratios obtained were: E/S = 0.224 = 43.67/194.97 (periodic environment); E/S = 0.224 = 43.45/193.87 (uniform-stochastic environment); E/S = 0.224 = 42.97/191.38 (Gaussian-stochastic environment). **(d)** The contribution of each variance to the overall sum varies as Z is relaxed (vertical dashed line corresponds to the optimal response Z = 1.25). **(e)** Variation of H = ln[*det*($\Sigma$)] and Cost/Score ratio in function of the constraint Z (scales of vertical axis are arbitrary). Maximum of H meets the minimal *E/S* ratio (0.224) at Z = 1.25 (vertical line).

Figure 6e depicts the effect of relaxing the constraint (8) on H. The maximum (H ≈ 2.28, periodic environment; H ≈ 1.77, stochastic (uniform) environment; H ≈ 4.18, stochastic (Gauss) environment) meets the minimal cost/score ratio (*E/S* ≈ 0.224). Numerical calculations gave identical results (Fig. SD4). The corresponding values of *E* and *S* form a compromise between the extreme values (Fig. SD2). The best combinations of {$\alpha_1, \alpha_2, \alpha_3$} parameters are respectively, {0.151, 0.374, 0.724}, {0.150, 0.373, 0.726} and {0.143, 0.364, 0.743} for the periodic and stochastic environments. These combinations should warrant the organism to produce the best response in fluctuating environments at the lowest possible cost of energy (fig 6a, 6b and 6c). Figure 6d shows clearly that the relative contributions of the ($\alpha_i$)s' do not evolve linearly with Z (see also Fig. SD3). The variation depends on the shape parameter $\beta_i$ of the $y_i$ responses. When the system is highly constrained, the gene *G2* ($\beta = 1.8$) is the most solicited. Next, its contribution diminishes in parallel with *G1* ($\beta = 2$) while the contribution of *G3* ($\beta = 1.0$) increases when Z is relaxed. The variations of weights both in costs and scores ($\gamma_i$, $\tau_i$, $\delta_i$, $\xi_i$) should likely modify this pattern.

## 4. Discussion & Conclusion

### 4.1. Optimal plasticity, optimal energy

Results show that in the same time the numerical model converges to the best scores, the determinant of the dispersion matrix diminishes toward a residual value, which is of comparable magnitude (H = 4.61) to the one obtained with the analytical model using a Gaussian-stochastic environment (H= 4.18). However, these similar outcomes cannot lead to clear and definite conclusions. Simply, we notice that in both cases, it persists (even after the convergence of the systems) a non negligible variance. It results from the latter consideration that the system never converges to a unique solution but to a set of solutions, each of them forming the best combination of the three variances associated to genes to give the best adaptive response to a particular environmental fluctuation in time. Residual value of H can thus be understood as a description of the optimal variance for best scores at lowest cost in the context of stochastic, but limited, environmental fluctuations.

Since the "niche construction" process (Odling-Smee et al., 2003) refers to the ability of an organism to alter its environment, the plasticity of characters seems to be particularly involved in the modifications of habitats:
1.  Optimal phenotypic plasticity allows providing the best response and consequently performing a *maximal direct work on the environment* (e.g., seed dispersal or germination timing for plants, foraging activity or habitat choice for animals...)
2.  The minimization of the cost/benefit ratio allows saving a certain amount of energy (compared to a non-optimized organism) which can be reallocated to other physiological compartments resulting in an *indirect work on the environment*.
Hence, according to the definition of exergy, one can assume that the optimum of phenotypic plasticity coincides with the optimum of exergy the organism can own (Jørgensen and Svirezhev, 2004) with respect to (i) the environmental signal the organism experiences and (ii) the genetic pathway involved in the response.

Obviously, the E/S value obtained is a theoretical optimum of an ideal organism. However, only few individuals of populations are likely to respond optimally. For several reasons: intra-population genetic variation, but also individuals are not in the same physiological state at a given time, to respond to environmental stimuli and therefore to devote the energy required for an optimal response. Hence, when an organism (or a population) has a very different plasticity than predicted by the H value, an accurate review of organisms living conditions must be made to determine which processes are acting to maintain them far from this value. Lastly, plasticity limits can probably lay in functional and internal constraints that reduce the benefit of plasticity in comparison with a perfect and optimal plasticity (see Gabriel *et al.*, 2005 for a discussion about the plasticity limits, and van Kleunen & Fisher (2005) for an analysis in the context of plants plasticity).

The H value calculated is used to describe the volume in which the organism is likely to deploy its activity to ensure survival, at the lowest cost, in the limits of environmental fluctuations it encounters. In other words, the H value describes the activity (a state space) of the organism, under survival constraints, and provides a metric for the dispersion of the genes' responses, taken as random variables. While many aspects of gene expression require an accurate knowledge of molecular mechanisms, the distribution of responses of genes is largely explained by a simple model based on statistics rather than a theory of molecular biology.

## 4.2. Plasticity and environmental variations

The accurate knowledge of functional response equations of genes involved in a phenotype and the associated costs should allow, applying this method, to predict –ad minima– the environmental fluctuations to which may successfully face the organisms. This feature could help in predicting the potentiality of species to become invasive when artificially transplanted into exogenous ecosystems or, on the contrary, to fail to survive (Donohue et al., 2001). In addition, according to Kaneko (2009) who found a theoretic positive correlation between genetic variance and phenotypic variance over many genes, one can find through the results presented here a basis to better understand the sustainability of such a level of variation, in spite of the natural selection that should act in direction of a reduced plasticity in the context of fixed environments. Thus, according to van Kleunen & Fisher (2005) one can make the conjecture that species confronting harsh and monotonous factors close to their extremum (arid areas, cold climates, salted soils…) or mesic but stable factors (tropical forests: Figueiredo-Goulart et al., 2006), should be less plastic than vicariant species accustomed to environments that are heterogeneous in space and time (Ellers and van Alphen, 1997; Snell-Rood et al., 2010; Tieleman, 2003).

## 4.3. Environmental variations as a source of genetic diversity

The results obtained here provide a restricted set of solutions close to the absolute optimum (in terms of cost and scores). This is clearly visible on fig. 6e, where acceptable solutions range approximately from *Z = 1* to *Z = 1.4*. The numerical model does not contradict this observation since scores (= fitness) after convergence of the algorithm exhibit a residual variance (*SD = 3.88*). Consequently, individuals can adopt any one of nearly equiprobable solutions. Besides, it can be inferred that some small differences between the genes' responses, or transitory variations of energy availability (Z), will be of feeble consequence on their fitness (Fig. SD2). Thus, even among individuals exhibiting a quasi-optimal phenotypic plasticity with respect to their environment, there might still persist a small, but hardly reducible, genetic variance, keeping an open door for further modifications of the genes' apparatus. This is an important point, since, "…higher [genetic] biodiversity means that a wider spectrum of properties is available for survival under changing conditions" (Jørgensen & Svirezhev, 2004). Hence, we agree with Kaneko (2009) and Svanbäck et al.(2009) who advocate for environmental variations (among other processes) as a fundamental root of evolvability.

In a same manner, Lande (2009) using a quantitative genetic model of plasticity found that major changes in the environment should first induce a substantial increase in genetic variance followed by a second phase where "the expected mean phenotype undergoes the last small fraction of the adaptation to the new expected optimum". However, the author also indicates that on a long time scale, "stabilizing selection around the expected optimum in the new average environment would re-canalize the genetic variance". But, this effect should remain a slow process. It results from these few considerations that a substantial part of genetic diversity could probably be hidden by an apparent homogeneity of the phenotypic plasticity in an ever changing world.

Lastly, we have to underline that further theoretical or empirical investigations on phenotypic plasticity should pay greater attention to the effects of both grain of environmental variation (coarse to fine-grained) and intensity of selection. Particularly, the role of these two interacting constraints on plasticity versus specialization is still in debate (Snell-Rood et al., 2010; Merilä et al. 2004) and its elucidation should be of greatest importance to clarify cost, limits, dynamics and properties of plasticity.

## 4.4. Alternative method

An alternative to the maximization of the determinant of the covariance matrix (section 3) could have been the use of a Structurally Dynamic Model (SDM). The latter is a heuristic method recently applied with success to more than 20 ecological problems (particularly to the process of lake eutophication (Zhang et al, 2010)). In SDM, parameters of functions vary continuously in order to reproduce the adaptation of the system to forcing variables that are steadily moving through time. These changes are obtained by means of the optimisation of a goal function. In most cases, the goal function used is, among others, the eco-exergy. The main idea of SDM is to find, at each step of the simulation, the set of parameters "that is better fitted for the prevailing conditions of the ecosystem" (Jørgensen, 2009).

Obviously, there are many similarities between the present work and the adaptive programming way of SDM. However, we notice that:

1. The present work mainly lies on statistics (we fitted some functional responses) rather than on detailed and realistic mechanistic relationships between the components of the system.
2. The optimization procedure we used is deterministic (Lagrange's constrained optimization method) and does not require prohibitive computing time.
3. The method presented here is unable to reproduce any dynamics and to produce prognoses such as SDMs can do. The method can only predict what should be the ideal parameters of an ideal system experiencing a fluctuant environment, and the distribution of the responses of its components.

Finally, it is clear that the choice of one of the two methods will depend on several considerations. A high complexity and an accurate knowledge of the main relationships between elements composing the system, objectives of the model such as shift of species composition and/or adaptation through time, the prognoses of the impact of environment modifications and the ability to elaborate a goal function should lead the modeller to the SDM choice.

## 4.5. Phenotypic plasticity and the MaxEnt algorithm

The most simple and convenient method for finding distribution functions, about which few is known, in statistical physics and information theory was proposed by E. T. Jaynes (1957a, 1957b) (see Fig. SD5 for a succinct description of the method and Martyushev and Seleznev (2006) for a complete review about Maximum Entropy Production Principle (MEPP) and its substantiation by Jaynes' Maximum Entropy Principle (MaxEnt)). Recently, the MaxEnt has been used to explain ecological patterns (Dewar and Porté, 2008; Harte et al., 2008; Azale et al., 2010; Banavar et al., 2010), biodiversity (Shipley et al., 2006) and foodweb distribution (Williams, 2009). However, some limitations in the application of the method to ecology have been pointed out (Banavar et al., 2010; Haegeman et al., 2009).

In the present case, the problem consists in finding the unknown distribution of $\{\alpha_1, \alpha_2, \alpha_3\}$ about which we only know the energetic cost this distribution induces. Let us suppose that the genes' response is multinormal. In such conditions, H should be maximized, with respect to the distributions of the *3* genes' responses given by (4), satisfying the constraint (7), where H (expressed in nats) is the differential entropy (Cover and Thomas, 2006) of the three genes (see below, Eq. (12)).

A brief description of the MaxEnt algorithm is given in supplementary information. Entropy of multivariate distributions has not been the subject of many works to date. A multivariate normal distribution admits two parameters $\mu$ (a vector), which defines its centre and a positive-definite symmetric matrix $\mathbf{\Sigma}$, which is the dispersion matrix of the random vector $\mathbf{X}$ ( $\mathbf{X} \in \Re^3$ in this case). The entropy of a multivariate distribution has the property:

$$H(\mathbf{X}) \leq \frac{1}{2}\ln((2\pi e)^n |\Sigma|), \qquad (12)$$

with equality iff $\mathbf{X} \sim \mathcal{N}(0, K)$, where $n = 3$ is the dimension of the distribution and $K = E\mathbf{X}\mathbf{X}^t$.

Obviously, there are some strong similarities between the calculation of $H = \ln|\Sigma|$ we made and the differential entropy (Eq. (12)). In fact, both calculations lead to identical results under the assumption that responses are multinormal distributions $p(S_i)$, which is, unfortunately, not the case since environment fluctuations $g(t)$ are not Gaussian (excepted in one case) and genes' responses induced by the signal are calculated through saturation functions.

This application can be considered as a transposition of the MaxEnt idea to a specific case, rather than a rigorous and standard application of the method. For two reasons at least: 1. We did not compute a true entropy since the dispersion matrix $\Sigma$ is obtained from the gene responses and not from probability distributions. 2. Doing so, one does not make any assumption about the distributions $p(S_i)$. Indeed, in such case, it is only supposed that the desired maximum entropy should be inferior to the maximum entropy of a multinormal-zero-mean distribution, whatever the distributions under study.

The results presented here, somehow bode a new way of applying the MaxEnt, with some restrictions (concave functions) but theoretical underpinnings of such a scheme require now further developments.

**Acknowledgments**. We would like to warmly thank Drs. René Feyereisen and Eric Wajnberg (INRA, Sophia-Antipolis), Dr James Nutaro (Oak Ridge National Laboratory, TN), and Prof. R. Dewar (Australian National University of Canberra) for fructuous discussions and suggestions about this paper. Simulations were performed thanks to the free softwares: IDE cross-platform Code::Blocks, the C++ Borland™ 5.5 compiler, the MIT's Galib library (Wall, 2007) and the environment for statistical computing R.

**REFERENCES**

Auld JR, Agrawal AA, Relyea RA. Re-evaluating the costs and limits of adaptive phenotypic plasticity, Proc. Royal. Soc. Belgium 2010; 277:503-511.

Azaele S, Muneepeerakul R, Rinaldo A, Rodriguez-Iturbe I. Inferring plant ecosystem organization from species occurrences, J. Theor. Biol. 2010; 262(2):323-329.

Banavar JR, Maritan A, Volkov I. Applications of the principle of maximum entropy: from physics to ecology, J Phys-Condens Mat 2010; 22(6):63–101.

Charnov EL, Optimal foraging: the marginal value theorem, Theor. Pop. Biol. 1976; 9:129–136.

Coquillard P, Muzy A, Wajnberg E. In: A. Muzy, D. Hill and B.P. Zeigler editors. Complex Systems: Activity-Based Modelling and Simulation, Clermont-Ferrand, Fr: UBP Press; 2010, p. 129–146.

Cover TM, Thomas JA. Elements of information theory, 2nd ed. Hoboken: Wiley-Interscience; 2006.

De Jong G, Phenotypic plasticity as a product of selection in a variable environment, Am. Nat. 1995; 145:493-512.

Dewar RC, Porté A. Statistical mechanics unifies different ecological patterns, J. Theor. Biol. 2008; 251(3):389–403.

DeWitt TJ, Sih A, Wilson DS. Costs and limits of phenotypic plasticity. Trends Ecol. Evol. 1998; 13:77–81.

DeWitt TJ. Costs and limits of phenotypic plasticity: test with predator-induced morphology and life history in a freshwater snail J. Evol. Biol. 1998; 11:465–480.

Dingemanse NJ, Kazem AJN, Réale D, Wright J. Behavioural reaction norm: animal personality meets individual plasticity. Trends Ecol. Evol. 2009; 25(2):81–89.

Donohue K, Pyle EH, Messiqua D, Heschel MS, Schmitt J. Adaptive divergence in plasticity in natural populations of Impatiens capensis and its consequences for performance in novel habitats, Evolution 2001; 55(4):692–702.

Donohue K. « Niche construction through phenological plasticity: life history dynamics and ecological consequences ». New Phytol. 2005; 166(1):83-92.

Ellers J, van Alphen JJM. Life history evolution in *Asobara tabida*: plasticity in allocation of fat reserves to survival and reproduction. J. Evol. Biol. 1997; 10:771–785.

Figueiredo-Goulart M, Pres Lemos Filho J, M.B. Lovato MB. Variability in Fruit and Seed Morphology Among and Within Populations of *Plathymenia* (Leguminosae - Mimosoideae) in Areas of the Cerrado, the Atlantic Forest, and Transitional Sites, Plant. Biol.2006; 8:112–119.

Gabriel W, Luttbeg B, Sih A, Tollrian R. Environmental tolerance, heterogeneity, and the evolution of reversible plastic responses, Am. Nat. 2005; 166:339–353.

Haegeman B, Loreau M. Trivial and non-trivial applications of entropy maximization in ecology: a reply to Shipley, Oikos 2009; 118(8):1270-1278.

Harte J, Zillio T, Conlisk E, Smith AB. Maximum entropy and the state-variable approach to macroecology, Ecology 2008; 89(10):2700-2711.

Jaynes ET, Information Theory and Statistical Mechanics II, Phys. Rev. 1957; 108-171.

Jaynes ET, Information Theory and Statistical Mechanics, Phys. Rev. 1957; 106-620.

Jørgensen SE, Svirezhev YM. Towards a thermodynamics theory for ecological systems. Amsterdam: Elsevier; 2004.

Jørgensen SE, in J. Devillers (ed.), Ecotoxicology Modeling (Emerging Topics in Ecotoxicology). The Application of Structurally Dynamic Models in Ecology and Ecotoxicology, Springer-Verlag, p 377-392, 2009.

Kaneko K. Relationship among phenotypic plasticity, phenotypic fluctuations, robustness and evolvability; Waddington's legacy revisited under the spirit of Einstein, J. Biosciences 2009; 34(4):529-542.

Lande R. Adaptation to an extraordinary environment by evolution of phenotypic and genetic assimilation, J. Evol. Biol.2009; 22:1435-1446.

Lynch M, Walsh B. Genetics and Analysis of Quantitative Traits. Sunderland MA: Sinauer Associates; 1998.

Martyushev L, Seleznev V. Maximum entropy production principle in physics, chemistry and biology, Phys. Rep. 2006; 426(1):1-45.

McNamara JM, Houston AI. Memory and the efficient use of information, J. Theor. Biol. 1987; 125:385-395.

Merilä J, Laurila A, Lindgren B. Variation in the degree and costs of adaptive phenotypic plasticity among *Rana temporaria* populations, J. Ecol. Evol. 2004; 17:1132-1140.

Miner, Benjamin G. et al. « Ecological consequences of phenotypic plasticity ». Trends Ecol. Evol. 2005; 20(12): 685-692.

Mouritsen KN, Poulin R. Parasites boosts biodiversity and changes animal community structure by trait-mediated indirect effects. Oikos 2005; 108:344–350.

Odling-Smee FJ, Laland, Marcus KN, Feldman W. Niche Construction: The Neglected Process in Evolution. Princeton, New Jesey: Princeton University Press.

Ramos-Jiliberto R, Mena-Lorca J, Flores JD, Morales-Alvarez W. Role of inducible defences in the stability of a tritrophic system. Ecol. Complex. 2008; 5.2:183-192.

Ramos-Jiliberto R. « Population dynamics of prey exhibiting inducible defences: the role of associated costs and density-dependence ». Theor. Popul. Biol. 2003; 64(2): 221-231.

Shipley B, Vile D, Garnier E. From plant traits to plant communities: a statistical mechanistic approach to biodiversity. Science 2006; 314:812-814.

Snell-Rood EC, Van Dyken JD, Cruickshank T, Wade MJ, Moczek AP. Toward a population genetic framework of developmental evolution: the costs, limits and consequences of phenotypic plasticity, BioEssays 2010; 32:71-81.

Svanbäck R, Pineda-Krch M, Doebeli M. Fluctuating population dynamics promotes the evolution of phenotypic plasticity, Am. Nat. 2009; 174(2):176-189.

Tieleman BI, Williams JB, Bushur ME, Brown CR. Phenotypic variation of larks along an aridity gradient: are desert birds more flexible? Ecology 2003; 84:1800-1815.

van Kleunen M, Fisher M. Constraints on the evolution of adaptive phenotypic plasticity in plants, New Phytol. 2005; 166:49–60.

Via S, Gomulkiewicz R, De Jong G, Scheiner SM, Schlichting SM, Van Tienderen PH. Adaptive phenotypic plasticity: consensus and controversy. Trends Ecol. Evol. 1995; 10:212-217.

Wall M, GAlib 2.4.7, 1995-2007, MIT, http://lancet.mit.edu/ga

Wajnberg E, Coquillard P, Vet LEM, Hoffmeister T. Optimal resource allocation to survival and reproduction in parasitic wasps foraging in fragmented habitats. PlosOne 2012; (In press).

Weinig C. Plasticity versus canalization: population differences in the timing of shade-avoidance responses, Evolution 2000; 54(2):441-451.

Whitley D, The GENITOR algorithm and selective pressure: why rank-based allocation of reproductive trials is best, Proceedings 3rd International Conference on Genetic Algorithms, eds D. Schaffer (Morgan Kaufmann), 1989, p. 116-121.

Williams RJ. Simple MaxEnt models explain food web degree distributions, Theor. Ecol. 2009; 3(1):45-52.

Zhang J, Gurkan Z, Jørgensen SE. Application of eco-exergy for assessment of ecosystem health and development of structurally dynamic models, Ecol. Model.2010; 221:693-702.